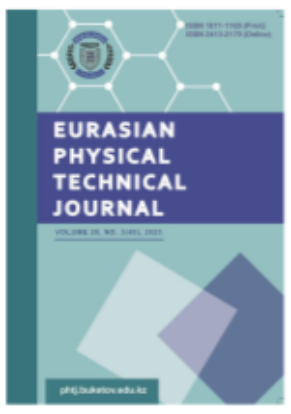

 

# COULD GLOBULAR CLUSTERS IN THE VICINITY OF THE GALACTIC CENTRE HAVE CONTRIBUTED STARS TO THE NUCLEAR STAR CLUSTER IN THE PAST?

Kalambay M.[1,2]*, Kuvatova D.[1,3], Ishchenko M.[4,5,1], Berczik P.[1,4], Omarov Ch.[1], Kusakin A.[1]

[1] Fesenkov Astrophysical Institute, Observatory 23, Almaty, Kazakhstan
[2] K. Zhubanov Aktobe Regional University, Aktobe, Kazakhstan
[3] Al-Farabi Kazakh National University, Almaty, Kazakhstan
[4] Main Astronomical Observatory, National Academy of Sciences of Ukraine, Kyiv, Ukraine
[5] National university of life and environmental sciences of Ukraine, Kyiv, Ukraine
*Corresponding author: kalambay@fai.kz

***Abstract.*** *The results of direct N-body simulations of three Milky Way bulge globular clusters – Terzan 2, Terzan 4, and Terzan 5 – evolving in a time-varying galactic potential derived from the IllustrisTNG-100 cosmological database over 8 Gyr are presented. Incorporating full stellar evolution and compact remnant formation, we quantify the probability and efficiency of stellar accretion onto the Galactic nuclear star cluster (NSC) during close pericentre passages of these three clusters. To parametrize the effect of dynamical friction in the dense central environment, we introduce a velocity reduction factor ($V_{red}$) ranging from 1.00 to 0.10. It was found that all three clusters contribute stars to the NSC even in the absence of velocity reduction, with a combined bound stellar count of ~730 stars and a total accreted mass of ~278 $M_\odot$ at $V_{red}$ = 1.00, growing to ~5420 stars and ~1175 $M_\odot$ at $V_{red}$ = 0.10. Terzan 4 is the dominant contributor across all velocity reduction factors, reflecting its more frequent and closer pericentre passages near the Galactic centre. The mass distribution of accreted stars reveals a bimodal structure: a low-mass component below ~1 $M_\odot$ whose capture is independent of $V_{red}$, and a prominent high-mass tail extending to ~20 $M_\odot$ corresponding to neutron stars and black holes, which grows with decreasing $V_{red}$. Notably, Terzan 4 is the only cluster for which white dwarfs are also captured by the NSC at every $V_{red}$ value. Our results suggest that the Terzan family represents a more significant source of compact remnants for the Galactic NSC than previously studied clusters, while confirming that tidal stripping of surviving globular clusters alone cannot account for the total NSC mass.*

## 1. Introduction

Nuclear star clusters (NSCs) are among the densest and most massive stellar systems and are found in the centers of most late-type galaxies [1]. In the case of the Milky Way (MW), the mass of the NSC is estimated to be ~107 $M_\odot$ [2]. Its stellar population includes both old and relatively young components, indicating a complex formation history [3]. One of the proposed formation mechanisms for NSCs is the migration and subsequent dissolution of globular clusters (GCs) due to dynamical friction, leading to their accretion onto the Galactic Centre (GalC) [4, 5]. Numerical models show that this process can proceed efficiently on cosmological timescales [6]. However, a number of studies suggest that the contribution of this mechanism is likely limited and can account for only a fraction of the NSC mass, while the remaining portion is formed through in situ star formation or gas accretion [7, 8]. An additional growth channel for NSCs can be associated not with the complete disruption of clusters, but with partial stellar mass loss during close passages through the GalC. In this case, even surviving clusters may contribute to the stellar population of the NSC [9]. It should be noted that the efficiency of this mechanism strongly depends on the orbital parameters of the clusters, their internal structure, and their evolutionary history [10].

In our previous works [6, 11], we investigated the dynamical evolution of six MW GCs (NGC 6401, NGC 6642, Palomar 6, NGC 6681, NGC 6981, and HP 1) selected based on a high probability of close passages to the GalC, using direct N-body simulations in a time-dependent potential derived from the cosmological database IllustrisTNG-100 [12, 13]. We showed that most of these clusters, which are predominantly bulge clusters, lose up to 70–90% of their mass over ~8 Gyr. A significant fraction of the stripped stars may potentially contribute to the formation of the NSC. However, only one GC - NGC 6642 exhibits a contribution to the NSC (primarily stars with masses < 1 $M_\odot$) of less than 0.1% of the current NSC mass, highlighting the strong dependence of the process on specific GC's orbital and structural properties.

At the same time, several bulge GCs, including Terzan 2, Terzan 4, and Terzan 5, were not included in the previous analysis, despite their proximity to the GalC and their potentially important dynamical role. These objects are of particular interest because, in addition to their close proximity to the GalC, they show signs of strong gravitational interactions with other clusters [14], which may significantly affect both their internal evolution and orbital dynamics. Terzan 5 occupies a special place, as its nature remains under debate. Unlike typical GCs, it exhibits a complex chemical and stellar structure, which has led to hypotheses that it may be the remnant of a protogalactic fragment or the product of mergers between multiple systems [15, 16]. This makes it a potentially non-standard source of stars for the NSC and warrants separate consideration. These three clusters were selected as a direct continuation of our previous study [6], in which we examined stellar accretion from clusters with close pericentre passages near the Galactic Centre but with apocentres located far from it (4–20 kpc). In the present work, we extend this analysis to clusters that evolve entirely within the central region of the Galaxy: Terzan 2, Terzan 4, and Terzan 5 all have apocentric distances of only 1–2 kpc, making them a natural sample for probing the accretion efficiency of clusters permanently confined to the Galactic Centre environment.

In this work, we focus on these three GCs, Terzan 2, Terzan 4, and Terzan 5, to assess their possible contribution to the formation and growth of the MW NSC. To this end, we perform direct N-body simulations of their evolution in a time-dependent galactic potential, including stellar evolution and the formation of compact remnants. This approach allows us to quantitatively evaluate the transfer of stars into the central region and its dependence on the properties of individual systems.

## 2. Initial conditions for physical models, modelling and mass loss

The initial conditions for each of the three selected Terzan-family clusters were chosen individually in order to reproduce the evolution of their masses and spatial characteristics. The initial stellar masses in each cluster were sampled from the Kroupa initial mass function [17] in the range 0.08-100 $M_\odot$. The individual metallicity of each cluster was taken into account in the subsequent stellar-evolution treatment [18]. To construct equilibrium initial models, we used King models [19] characterised by two free parameters: the half-mass radius $r_{hm}$ and the dimensionless concentration parameter $W_0$. Thus, for each cluster, three key initial parameters – the mass M, $r_{hm}$, and $W_0$ – were selected at t = –8 Gyr relative to the present day. Their values were chosen so that, by the present epoch, the model clusters reproduced the observed properties of their real counterparts. The details of the fitting procedure are given in our earlier work [6], whereas a more complete description of the numerical setup is presented in our recent article [20]. The initial masses, half-mass radii, King concentration parameters, and metallicities adopted for the Terzan 2, Terzan 4, and Terzan 5 models at (t = –8 Gyr) Gyr are summarised in Table 1.

In addition, a scaling approach in which one numerical particle represents ten real stars of the same type was used. A detailed justification and description of this method are provided in [21]. The full N-body simulations were carried out using the parallel high-order φ-GPU code[1] [22], into which a time-dependent MW-like galactic potential was incorporated.

**Table 1.** Initial physical characteristics at -8 Gyr for the Terzans.

| GC | M, $10^6$ $M_\odot$ | N | $r_{hm}$, pc | $W_0$ | $Z_m$, $10^{-3}$ |
|---|---|---|---|---|---|
| Ter2 | 1.1 | 1 916 827 | 5.0 | 9.0 | 3.17 |
| Ter4 | 1.3 | 2 265 341 | 5.0 | 9.0 | 0.50 |
| Ter5 | 2.3 | 4 007 900 | 5.0 | 3.0 | 3.17 |

A model with ID 411321 from the IllustrisTNG-100 cosmological database was aaccepted [12, 13], whose properties are close to those of the MW at zero redshift. In addition, a single-star evolution prescription [23] was implemented in the code. Together, these components allowed us to improve the physical realism of our simulations. The mass evolution of Terzan 2, Terzan 4, and Terzan 5 over 8 Gyr is shown in Fig. 1. Terzan 2 and Terzan 4 exhibit similar mass-loss histories, each losing approximately 75% of its initial mass by the present day, whereas Terzan 5 loses about 45%. In Fig. 2 we present the large-scale view of today's density distributions for Terzans family after the 8 Gyr of dynamical evolution, which was calculated based on Equation (1) from our previous article [6].

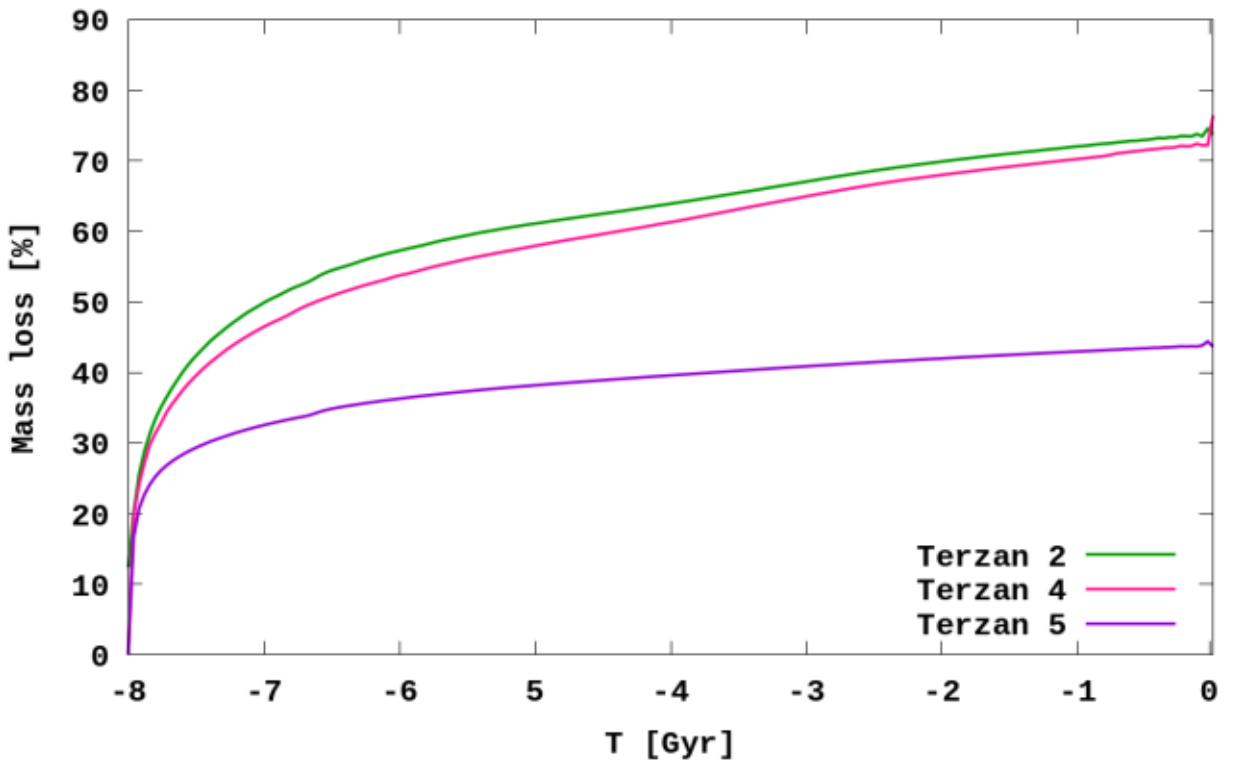

**Fig.1.** Evolution of the mass loss in the Terzan 2, 4 and 5 in percent due to stellar evolution and orbital type.

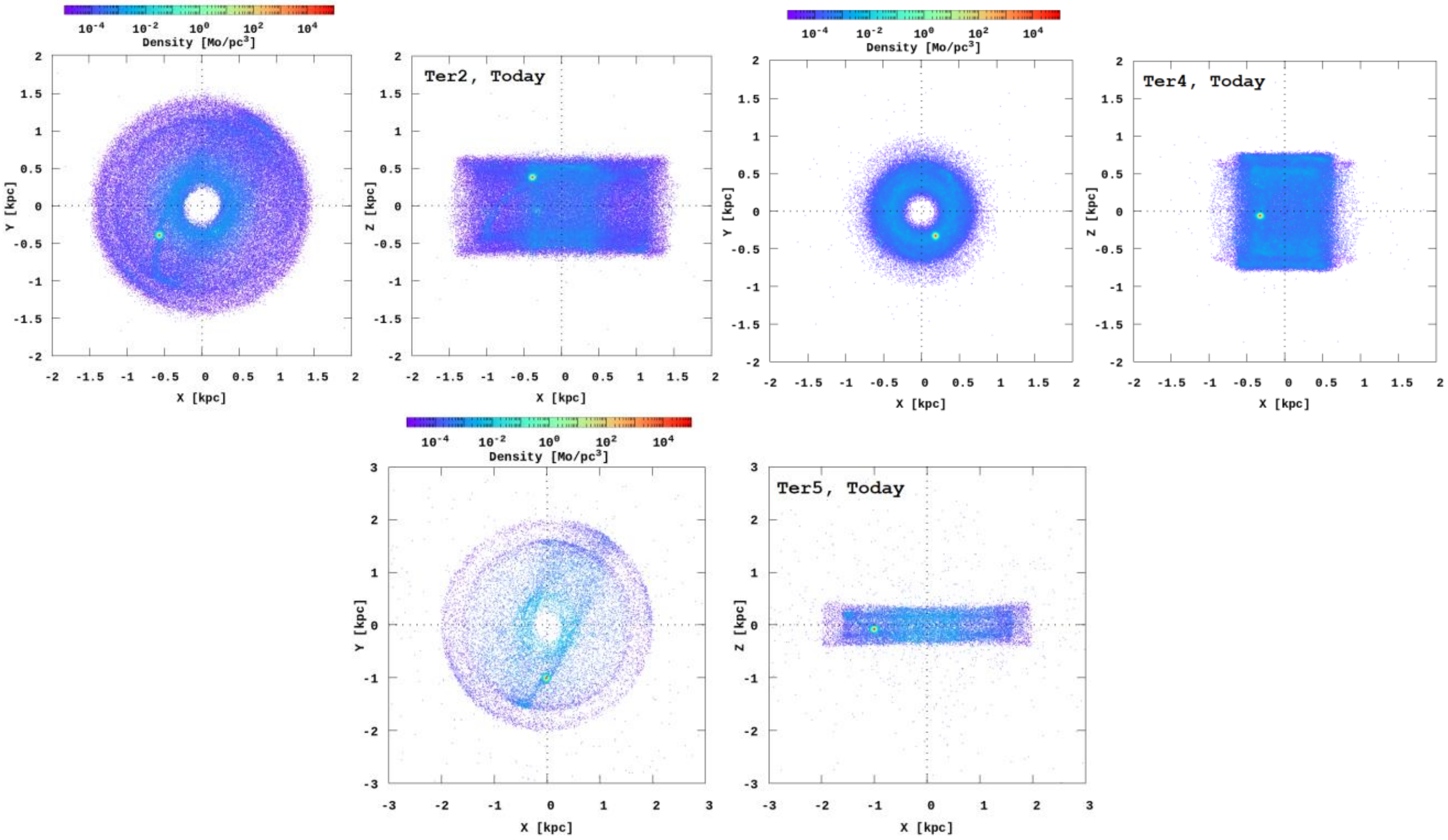

**Fig.2**. The stellar density distributions of Terzan 2, Terzan 4, and Terzan 5 after 8 Gyr of modelling is illustrated through two projections: (X, Y) and (X, Z).

It should be noted that based on applied Kroupa mass function for our stellar single generation prescription, we obtain the following total numbers of proto-stellar remnants — black holes (BH), neutron stars (NS), and white dwarfs (WD) — for each cluster: Terzan 2: BH = 3190, NS = 12030, WD = 172720; Terzan 4: BH = 3890, NS = 17070, WD = 218280; Terzan 5: BH = 5840, NS = 22630, WD = 321700. Due to the implementation of up-to-date natal kick prescription for BH and NS in our dynamical modelling, after 8 Gyr of evolution we obtain the next amount of the BH and NS which are still bound to their host GCs: Terzan 2: BH = 1780, NS = 840, WD = 93130; Terzan 4: BH = 2280, NS = 1460, WD = 101730; Terzan 5: BH = 3980, NS = 2930, WD = 315720. As expected, almost all of them are located in the central region of the GCs, less than GC's tidal radii.

## 3. Results and discussion

To evaluate the probability of stellar accumulation onto the Galactic NSC from Terzan 2, Terzan 4, and Terzan 5, we followed individual stellar trajectories throughout the full 8 Gyr of dynamical evolution and started to record each event in which a star entered the inner GalC zone within D < 100 pc. After that, we identified the closest approach (minimum distance) to the NSC and calculated the status of the star in relation to the NSC — whether it was bound or not. Following the approach of Ishchenko et al. (2024) [6], we introduced a relative velocity reduction factor $V_{red}$ to account for the effect of dynamical friction in the dense central environment, which is not explicitly modelled in our N-body simulations. The bound status of each star during its closest passage near the NSC potential was evaluated using the energy criterion:

$$E^{*}_{bound} = -\frac{(M_{NSC}+m_{*})}{R_{*}} + V_{red} \cdot 0.5 \cdot V_{*}^{2}, \quad (1)$$

where MNSC is the NSC mass, $R_*$ is the separation between the NSC and the star, and $V_*$ is the relative velocity between them. Stars with $E^{*}_{bound} < 0$ are considered bound to the NSC. This parameterisation allows us to explore a range of dynamical friction scenarios from the no-friction limit ($V_{red}$ = 1.00) to strong deceleration ($V_{red}$ = 0.10).

Fig. 3 shows the cumulative number distributions of NSC-bound stars as a function of their orbital parameters – pericentric distance $r_{per}$ and orbital eccentricity ecc for the combined Terzans. As $V_{red}$ decreases, the maximum pericentric distance of bound stars shifts to larger values, reflecting the artificial velocity reduction applied in the energy estimate (Eq. 1). The bound particles generally exhibit high orbital eccentricities (≳0.8), and consequently large semimajor axes.

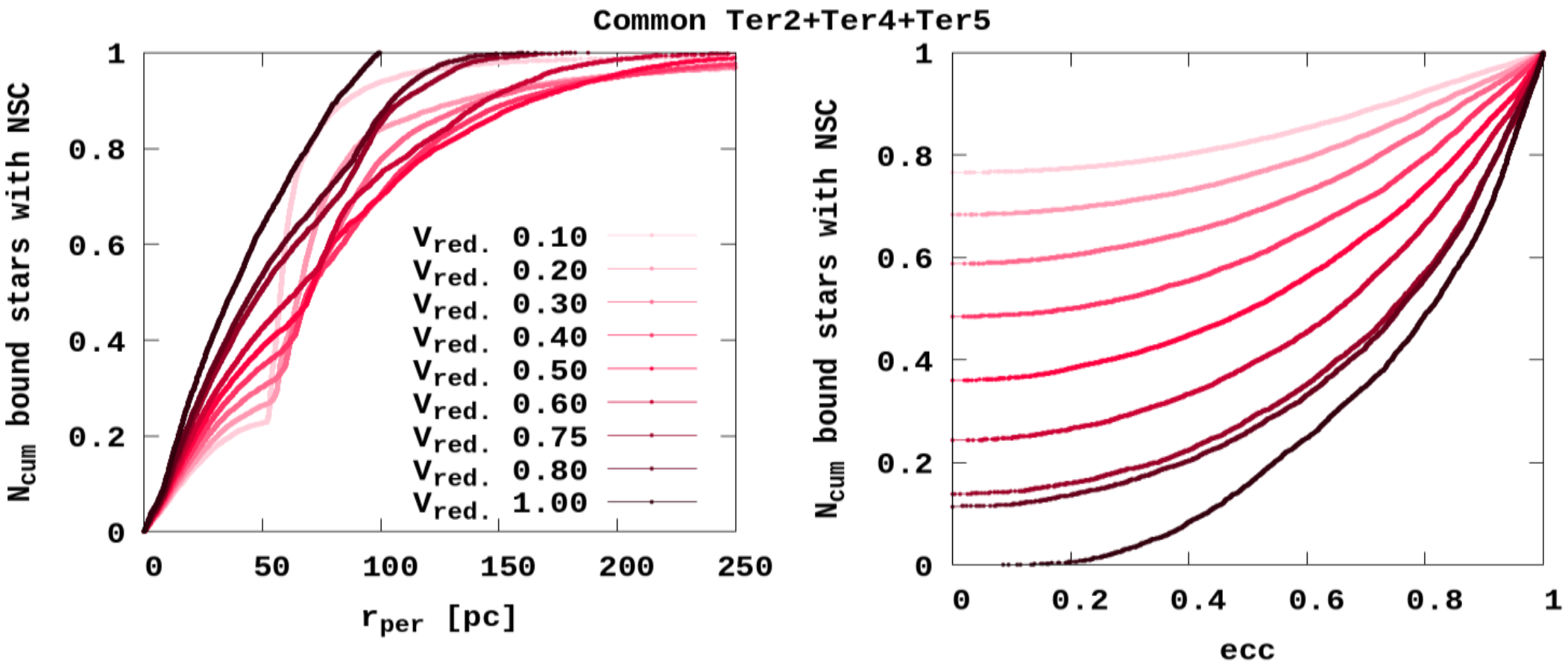

Fig. 3. Cumulative number distributions of particles bound to the NSC as a function of the orbital parameters — pericentric distance $r_{per}$ and orbital eccentricity ecc - shown for different velocity reduction factors for the three Terzan clusters.

The total number of bound stars and their accumulated masses for all $V_{red}$ values are summarised in Table 2. The total present-day masses of the three clusters are: Terzan 2 – 6.70×104 M⊙, Terzan 4 – $8.19\times10^{4}$ M⊙, and Terzan 5 – $1.25\times10^{5}$ M⊙, giving a combined mass of $2.74\times10^{5}$ M⊙.

**Table 2.** Total number of the bound particles with the NSC and their masses with different velocity reduction.

| $V_{red}$ | Ter2 | | Ter4 | | Ter5 | | Ter2+Ter4+Ter5 | |
|---|---|---|---|---|---|---|---|---|
| | M⊙, % | N, % | M⊙, % | N, % | M⊙, % | N, % | M⊙, % | N, % |
| $V_{red}$ 1.00 | 4.643e+01, 0.06926 | 130, 5.18 | 2.032e+02, 0.2481 | 470, 1.72 | 2.793e+01, 0.02234 | 130, 1.96 | 2.775e+02, 0.1013 | 730, 2.00 |
| $V_{red}$ 0.80 | 7.577e+01, 0.1130 | 220, 3.69 | 2.346e+02 0.2865 | 630, 1.81 | 3.917e+01, 0.03134 | 160, 1.65 | 3.496e+02, 0.1276 | 1010, 2.00 |
| $V_{red}$ 0.75 | 9.022e+01, 0.1346 | 250, 3.12 | 2.393e+02, 0.2921 | 670, 1.81 | 7.477e+01, 0.05983 | 220, 1.96 | 4.042e+02, 0.1476 | 1140, 2.03 |
| $V_{red}$ 0.60 | 1.199e+02, 0.1789 | 350, 2.71 | 2.838e+02, 0.3465 | 820, 1.86 | 1.453e+02, 0.1163 | 400, 2.17 | 5.490e+02, 0.2004 | 1570, 2.08 |
| $V_{red}$ 0.50 | 1.367e+02, 0.2039 | 430, 2.67 | 3.314e+02, 0.4047 | 970, 2.01 | 1.834e+02, 0.1467 | 480, 1.98 | 6.515e+02, 0.2379 | 1880, 2.12 |
| $V_{red}$ 0.40 | 1.490e+02, 0.2223 | 530, 2.75 | 3.682e+02, 0.4496 | 1130, 2.14 | 2.222e+02, 0.1778 | 580, 1.99 | 7.394e+02, 0.27 | 2240, 2.21 |
| $V_{red}$ 0.30 | 1.683e+02, 0.2510 | 690, 3.12 | 4.169e+02, 0.5091 | 1450, 2.5 | 2.381e+02, 0.1905 | 710, 2.22 | 8.232e+02, 0.3006 | 2850, 2.54 |
| $V_{red}$ 0.20 | 1.930e+02, 0.288 | 900, 3.51 | 4.918e+02, 0.6005 | 1860, 2.93 | 2.785e+02, 0.2229 | 1000, 2.69 | 9.633e+02, 0.3517 | 3760, 2.98 |
| $V_{red}$ 0.10 | 2.318e+02, 0.3458 | 1210, 4.38 | 6.017e+02, 0.7347 | 2710, 4.03 | 3.416e+02, 0.2733 | 1500, 3.71 | 11.75e+02, 0.4290 | 5420, 4.00 |

* M, % — accreted mass as a percentage of the cluster's total present-day mass; N, % — number of unique bound stars as a percentage of the total number of bound-star records (repeated close passages of the same star counted separately).

Even in the absence of velocity reduction ($V_{red}$ = 1.00), all three clusters contribute stars to the NSC: Terzan 2 contributes ~130 stars (~46 M⊙, 0.069% by mass), Terzan 4 ~470 stars (~203 M⊙, 0.25%), and Terzan 5 ~130 stars (~28 M⊙, 0.022%). The combined system delivers ~730 stars with a total mass of ~278 M⊙ at $V_{red}$ = 1.00. As $V_{red}$ decreases to 0.10, this grows to ~5420 stars and ~1175 M⊙ for the combined system, with Terzan 4 consistently providing the largest absolute contribution across all reduction factors.

Fig. 4 shows the cumulative stellar mass distributions as a function of individual stellar mass at the moment of closest approach to the GalC for $V_{red}$ = 1.00, 0.75, 0.50, and 0.20 (combined Ter2 + Ter4 + Ter5). The number of low-mass stars (below 1 M⊙) bound to the NSC remains constant across all four velocity reduction factors: 160 stars with a total mass of ~3.82 M⊙. This value is the same for all $V_{red}$, indicating that the capture of low-mass stars is entirely independent of velocity deceleration and is governed purely by the orbital dynamics during close pericentre passages. However, given that 160 stars represents a statistically small sample relative to the total bound population, this low-mass component is not shown in Fig. 4 to avoid misleading visual impressions at this scale; it is reported here in the text for completeness. The distributions in Fig. 4 therefore begin at ~1 M⊙ and are dominated by the high-mass tail corresponding to NSs and BHs, which grows progressively with decreasing $V_{red}$, reflecting the increasing efficiency of compact remnant capture under stronger velocity deceleration. This is consistent with the findings of Ishchenko et al. (2024) [6] for NGC 6642, where the vast majority of accreted objects are low-mass stars – in our case, however, the Terzan family shows a much more prominent compact remnant contribution even at realistic $V_{red}$ values.

The relative efficiency with which each cluster contributes stars and remnants to the NSC can be understood in terms of which stars actually reach the central region: these are predominantly stars that have become unbound from their host cluster and populate its tidal tails, rather than stars still gravitationally bound to the cluster core. Terzan 5 retains an unusually dense core even after 8 Gyr of evolution, owing to its compact initial half-mass radius (~2 pc) and high initial concentration parameter (W0 = 8; see Fig. 1), which limits the fraction of stars lost into tidal tails and therefore available for capture by the NSC — accounting for its comparatively low contribution relative to its total mass despite being the most massive of the three clusters.

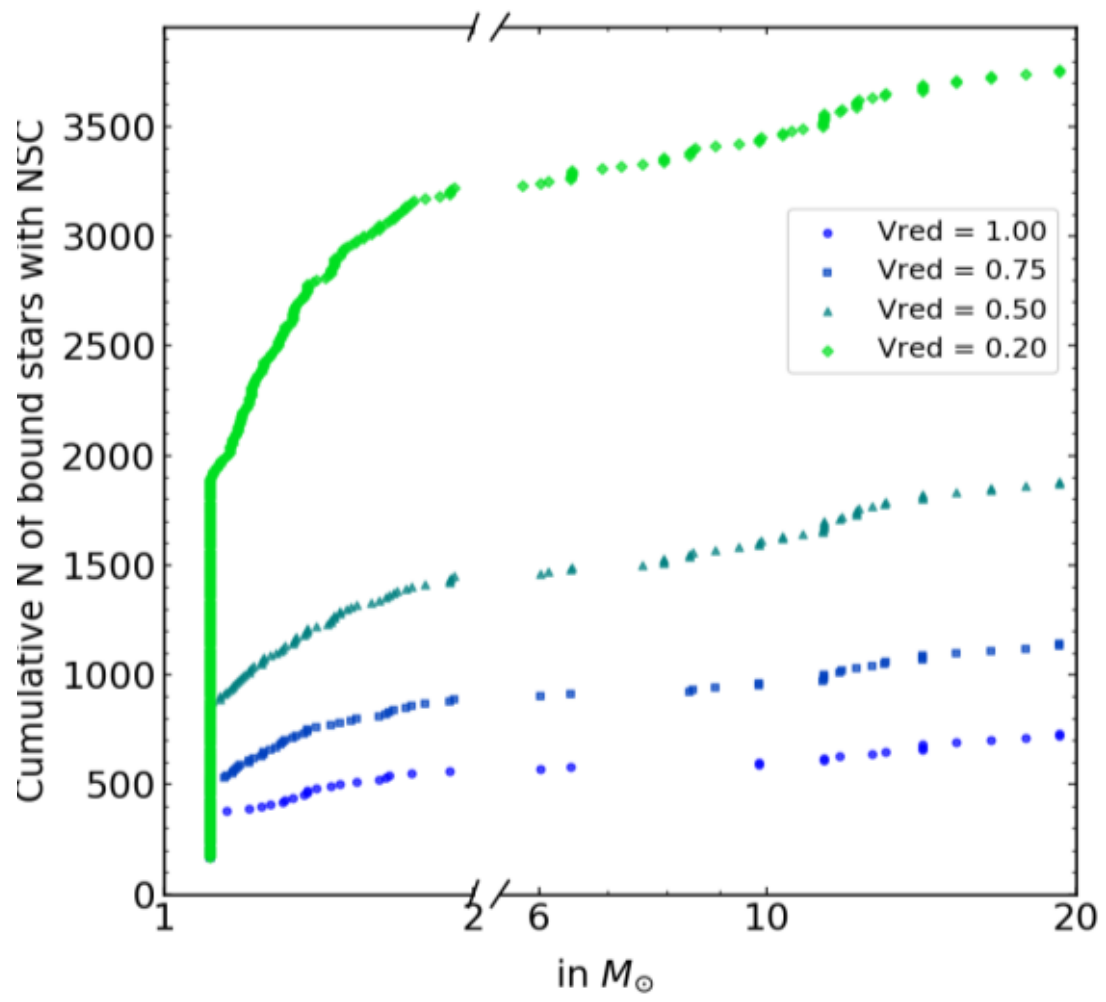

**Fig. 4.** Cumulative number of stars bound to the NSC as a function of individual stellar mass (in $M_\odot$) for Ter2 + Ter4 + Ter5 at the moment of closest approach to the GalC, shown for four velocity reduction factors ($V_{red}$ = 1.00, 0.75, 0.50, and 0.20). Note: the low-mass component (< 1 $M_\odot$) is omitted for clarity and reported separately in the text.

Terzan 2 and Terzan 4, by contrast, lose a comparable fraction of their mass into tidal tails over the same period; the larger accreted population from Terzan 4 therefore reflects its orbit, whose pericentre lies closer to the Galactic Centre, rather than a difference in mass-loss efficiency between the two clusters.

The fraction of compact stellar remnants – BHs and NSs – captured by the NSC is presented in Fig. 5, separately for each cluster and their combined contribution. These fractions increase substantially with decreasing $V_{red}$: at 0.10, Terzan 4 reaches 6.94% bound BHs and 13.42% bound NSs (270 BHs, 2290 NSs), the highest values in the sample. Terzan 2 reaches 2.82% bound BHs and 9.23% bound NSs (90 BHs, 1110 NSs), and Terzan 5 reaches 3.25% bound BHs and 5.79% bound NSs (190 BHs, 1310 NSs) at the same reduction factor.

Notably, Terzan 4 is the only cluster in our sample for which compact remnants of all three types – BHs, NSs, and WDs – are captured by the NSC. Specifically, 10 WDs are bound to the NSC from Terzan 4 consistently across all $V_{red}$ values, while no WDs are captured from Terzan 2 or Terzan 5 at any $V_{red}$. This can be attributed to the particularly dense and frequent dynamical interaction of Terzan 4 with the NSC, which makes it the most strongly coupled cluster to the GalC environment among the three. Despite the always-high natal kick velocities associated with NS formation, the number of NSC-bound NSs is comparable to – and in many cases exceeds – the number of bound BHs. This behaviour is consistent with findings in Ishchenko et al. (2024) [6] and reflects the large total number of NSs formed relative to BHs, which compensates for their lower retention probability. The contribution of the captured NSs to the NSC and their implications for the gamma-ray flux from the GalC are discussed in detail in Kalambay et al. (2026) [24].

In [20], the long-term dynamical evolution of Terzan 2, Terzan 4, and Terzan 5 was studied in detail through high-resolution direct N-body simulations over 8 Gyr. It was demonstrated that all three clusters undergo substantial mass loss: Terzan 2 and Terzan 4 each lose approximately 75% of their initial mass, while Terzan 5 loses about 45%. Multiple close encounters between the clusters were identified, with the most significant interactions occurring between the Ter2-Ter4 and Ter4-Ter5 pairs, in some cases with separations smaller than the sum of their tidal radii. Terzan 4 was found to be the most dynamically active cluster, driving triaxial deformations in Terzan 2 and Terzan 5 through repeated gravitational perturbations. In the present work, the analysis is extended by focusing on the second key aspect: the probability and efficiency of stellar and compact remnant accretion onto the Galactic NSC during close pericentre passages of these three clusters. These three clusters were selected as representatives of the GC population located closest to the GalC, making them natural candidates for NSC mass growth through tidal stripping. The contribution of each cluster to the stellar population and compact object content of the Galactic NSC was quantified using the velocity reduction factor $V_{red}$ to parametrise the effect of dynamical friction in the dense central environment.

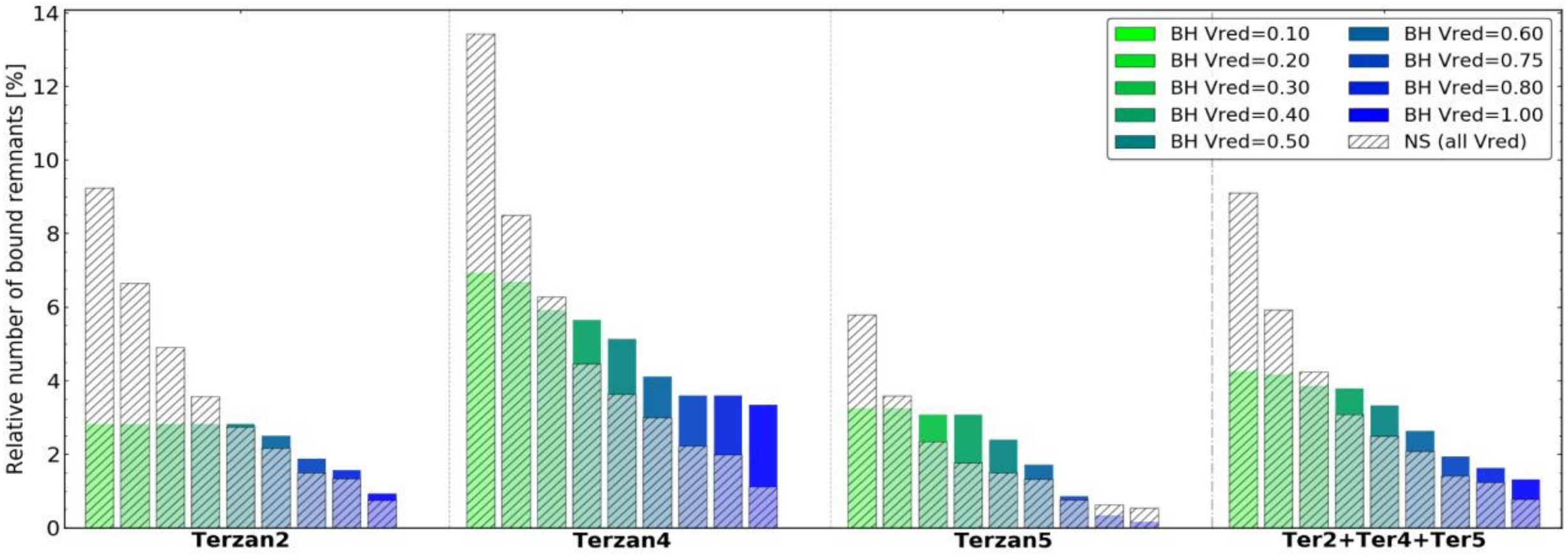

**Fig. 5**. Fraction of compact stellar remnants (BHs and NSs) bound to the NSC for Terzan 2, Terzan 4, Terzan 5, and their combined contribution (Ter2 + Ter4 + Ter5), shown across ten velocity reduction factors (Vred = 0.10–1.00); BHs and NSs are distinguished by different colours and symbols.

The mass distribution of the accreted stellar population reveals a clear bimodal structure (Fig. 4). The number of low-mass stars (below 1 M⊙) bound to the NSC remains constant across all four velocity reduction factors: 160 stars (scaled from 16 simulation particles) with a total mass of ~3.82 M⊙. This indicates that the capture of low-mass stars is independent of $V_{red}$ and is governed purely by orbital dynamics during close pericentre passages. Due to the small size of this low-mass sample relative to the total bound population, it is not shown in Fig. 4 but is reported here for completeness. The distributions in Fig. 4 are therefore dominated by the high-mass tail extending up to ~20 M⊙, corresponding to NSs and BHs, which grows progressively with decreasing $V_{red}$. This behaviour is qualitatively different from what was found for NGC 6642 in Ishchenko et al. (2024) [6], where the accreted population was dominated by low-mass stars with negligible compact remnant contribution at realistic $V_{red}$ values. The Terzan clusters, particularly Terzan 4, thus represent a potentially more significant source of compact remnants for the Galactic NSC.

Compact remnants are captured by the NSC across all $V_{red}$ values. At $V_{red}$ = 0.10, Terzan 4 contributes 270 BHs (6.94%) and 2290 NSs (13.42%), the highest fractions in the sample; Terzan 2 contributes 90 BHs (2.82%) and 1110 NSs (9.23%); and Terzan 5 contributes 190 BHs (3.25%) and 1310 NSs (5.79%). Additionally, Terzan 4 is the only cluster for which 10 WDs are captured by the NSC consistently at every $V_{red}$ value, attributable to its uniquely dense and frequent dynamical coupling with the GalC environment. Despite the always-high natal kick velocities associated with NS formation, the number of NSC-bound NSs is comparable to or exceeds that of BHs in all three clusters, owing to the much larger total number of NSs formed over the course of cluster evolution. This disproportionate remnant contribution may locally enhance the BH and NS content of the Galactic NSC relative to a standard stellar population. The implications of the captured NSs for the gamma-ray flux from the GalC are discussed in detail in Kalambay et al. (2026) [24].

## 4. Conclusion

*The main results of this study are as follows.* Terzan 2 provides a moderate but consistent contribution across all $V_{red}$ values, reaching ~1210 stars and ~232 M⊙ at $V_{red}$ = 0.10. While its absolute contribution is lower than that of Terzan 4, it remains significant across the full range of velocity reduction factors, reflecting its relatively compact orbit and frequent pericentre passages near the GalC. At $V_{red}$ = 1.00, Terzan 2 already delivers ~130 stars (~46 M⊙), confirming that stellar capture occurs even without any velocity deceleration. Only BHs and NSs are captured from Terzan 2; no WDs are bound to the NSC at any $V_{red}$ value.

Terzan 4 is the dominant contributor across all $V_{red}$ values, both in stellar number and total accreted mass, reaching ~2700 stars and ~602 M⊙ at $V_{red}$ = 0.10. Its higher contribution compared to the other two clusters reflects its orbital properties – in particular more frequent and closer pericentre passages near the GalC – and is further evidenced by the unique capture of compact remnants of all three types (BHs, NSs, and WDs) at every $V_{red}$ value, which is not observed for Terzan 2 or Terzan 5. Terzan 5 contributes ~1500 stars and ~342 $M_\odot$ at $V_{red}$ = 0.10, making it the second largest contributor by stellar number in the sample. Despite being the most massive cluster of the three — with a present-day mass of ~$1.25 \times 10^5$ $M_\odot$ — its contribution as a fraction of its total mass remains the smallest among the three clusters, reflecting the lower orbital coupling of Terzan 5 with the NSC compared to Terzan 4. At $V_{red}$ = 1.00, Terzan 5 contributes ~130 stars (~28 $M_\odot$), comparable in number to Terzan 2. Similar to Terzan 2, no WDs are captured from Terzan 5 at any $V_{red}$ value, and the captured remnant population consists exclusively of BHs and NSs.

Taken together, these results indicate that all three members of the Terzan family are capable of contributing stars and compact remnants to the Galactic NSC, with Terzan 4 being the most efficient contributor across all remnant types. The efficiency of this process depends critically on the degree of velocity deceleration experienced by stars during their closest passages through the NSC potential. A significant contribution even at $V_{red}$ = 1.00 distinguishes the Terzan clusters from several of the clusters studied in Ishchenko et al. (2024) [6], where NSC-bound stars only appeared at lower $V_{red}$ values, and underlines the importance of the specific orbital configurations of Terzan 2, Terzan 4, and Terzan 5.

An additional contribution may originate from clusters that were disrupted early in the Galaxy's history and are therefore absent from the present-day GC population. In work [25], we investigated this channel using 150 theoretical GC models evolved in a time-dependent Milky Way-like potential from 10 to 5 Gyr ago, including stellar evolution. Accretion onto the proto-NSC was most efficient during the first billion years of evolution, with the most compact models providing the largest total contribution. The accreted stellar population was dominated by low-mass stars (~ 0.33 $M_\odot$). The total accreted mass from these models reached approximately $4 \times 10^5$ $M_\odot$, corresponding to about 2% of the present-day NSC mass of $2 \times 10^7$ $M_\odot$) adopted in that study. An extended estimate incorporating higher-mass cluster models, under the assumption of the same frequency distribution as for the lower-mass population, yielded a contribution of approximately 6%. This estimate was interpreted as a conservative lower limit within the adopted modelling framework. These findings indicate that early cluster disruption can provide an additional contribution beyond that inferred from surviving GCs, although the populations considered in that study do not reproduce the entire present-day NSC mass.

The present results, together with those of Ishchenko et al. (2024) & Kuvatova et al. (2026) [6, 25], consistently show that individual surviving MW GCs do not provide substantial stellar accretion to the NSC under realistic dynamical friction conditions. Even in the most favourable cases – such as Terzan 4 and NGC 6642 – the accreted mass represents only a small fraction of the NSC total mass of ~$10^7$ $M_\odot$. This confirms that the inspiral and tidal stripping of surviving clusters alone cannot account for the formation of the Galactic NSC, and that other channels – such as the complete dissolution of now-defunct clusters or in situ star formation – must have played a dominant role in its assembly. Thus, summarizing the results obtained in this study and the two preceding papers, we can conclude that, based on the N-body modeling, the current mass of the NSC could not have formed solely through the accretion of stars from globular clusters—whether those observed today or those that may have been disrupted by now. Most likely, both mechanisms are involved in such a process, including formation of stars from gas that has previously migrated to the galactic center.

**Conflict of interest statement**
The authors declare that they have no conflict of interest in relation to this research, whether financial, personal, authorship or otherwise, that could affect the research and its results presented in this paper.

**CRediT author statement**
**M. Kalambay:** Formal analysis, Data curation, Writing – original draft. **D. Kuvatova:** Formal analysis, Data curation, Writing – original draft. **M. Ishchenko:** Conceptualization, Software, Investigation, Visualization, Writing – original draft. **P. Berczik:** Conceptualization, Methodology, Software, Supervision, Writing – review & editing. **Ch. Omarov:** Writing – review & editing. **A. Kusakin:** Writing – review & editing.The final manuscript was read and approved by all authors.

**Statement on the use of Artificial Intelligence.**
During the preparation of this manuscript, artificial intelligence tools were used solely for language editing and grammatical improvement.

**Data Availability Statement**
The data are available upon reasonable request from the authors.

**Funding**

This research was supported by the Ministry of Science and Higher Education of the Republic of Kazakhstan under Project No. AP26100669, "Galactic Archaeology of Gaia-Enceladus Globular Clusters as Indicators of the Formation History of the Milky Way Disk", and by the Aerospace Committee of the Ministry of Artificial Intelligence and Digital Development of the Republic of Kazakhstan under Program No. BR34836926, "Research and Monitoring of Near-Earth Space and Deep Space Based on the Development and Scaling of the Kazakhstan Network of Optical Telescopes".

**Acknowledgements**

Authors gratefully acknowledge the Polish high-performance computing infrastructure PLGrid (HPC Center: ACK Cyfronet AGH) for providing computer facilities and support within computational grant No. PLG/2026/019243.

---

## AUTHORS' INFORMATION

**Kalambay, Mukhagali** – PhD, researcher, Fesenkov Astrophysical Institute, Almaty, Kazakhstan; Associate professor, Heriot-Watt International Faculty, K. Zhubanov Aktobe Regional University, Aktobe, Kazakhstan. Scopus Author ID: 59365823900; ORCID: 0000-0002-0570-7270; kalambay@fai.kz

**Kuvatova, Dana** – PhD student, Al-Farabi Kazakh National University, Fesenkov Astrophysical Institute, Almaty, Kazakhstan; Scopus Author ID: 57980506300; ORCID: 0000-0002-5937-4985; kuvatova@fai.kz

**Ishchenko, Maryna** – Dr., Leading Researcher, Main Astronomical Observatory, National Academy of Sciences of Ukraine, Kyiv, Ukraine; Fesenkov Astrophysical Institute, Almaty, Kazakhstan; Scopus Author ID: 55348120800; ORCID: 0000-0002-6961-8170; marina@mao.kiev.ua

**Berczik, Peter** – Dr (Sci)., Leading Researcher, Fesenkov Astrophysical Institute, Almaty, Kazakhstan; Main Astronomical Observatory, NAS of Ukraine, Kyiv, Ukraine; Scopus Author ID: 8681718700; ORCID: 0000-0003-4176-152X; berczik@mao.kiev.ua

**Omarov, Chingis** – Dr., Professor, Fesenkov Astrophysical Institute, Almaty, Kazakhstan; Scopus Author ID: 8655858100; ORCID: 0000-0002-1672-894X; chingis.omarov@fai.kz

**Kusakin, Anatoly** – Dr., Leading Researcher, Fesenkov Astrophysical Institute, Almaty, Kazakhstan; Scopus Author ID: 6603251549; ORCID: 0000-0002-7756-546X; kusakin@fai.kz